\documentclass[11pt]{article}

\usepackage[margin=1in]{geometry}
\usepackage{graphicx}
\usepackage{amsmath}
\usepackage{amssymb}
\usepackage{hyperref}
\usepackage{enumitem}
\usepackage{booktabs}
\usepackage{float}
\usepackage{tikz}
\usetikzlibrary{positioning}
\usepackage{subcaption}
\usepackage{placeins}

\hypersetup{
    colorlinks=true,
    linkcolor=blue,
    citecolor=blue,
    urlcolor=blue
}

\title{A Modular Multi-Document Framework for Scientific Visualization and Simulation in Java}

\author{
D. Heddle \\
School of Engineering and Computing \\
Christopher Newport University \\
\texttt{david.heddle@cnu.edu}
}

\date{\today}

\begin{document}

\maketitle

\begin{abstract}
This paper presents the design and implementation of a modular multi-document interface (MDI) framework for scientific visualization and simulation in the Java Virtual Machine (JVM) ecosystem. The framework emphasizes architectural separation between visualization layers, simulation engines, and optional hardware-accelerated 3D rendering. 3D functionality is isolated into a separate module to prevent unnecessary dependency coupling in 2D-only applications.

We describe the core abstractions, threading model, simulation integration strategy, and dependency isolation approach. A case study involving a real-time 3D gas expansion simulation integrated with synchronized 2D entropy plotting demonstrates architectural cohesion. The framework is publicly available via Maven Central and targets long-lived scientific and engineering desktop applications.
\end{abstract}

\section{Introduction}

Scientific and engineering desktop applications frequently require architectural properties distinct from those emphasized in contemporary web and mobile development. These include:

\begin{itemize}
    \item Deterministic rendering behavior
    \item Long-running simulation support
    \item Offline deployment capability
    \item Stability across multiple Java releases
    \item Minimal runtime dependency complexity
\end{itemize}

While modern UI toolkits often prioritize declarative interfaces and rapid visual iteration, scientific desktop systems typically prioritize robustness and long-term maintainability. This paper describes the architectural considerations underlying a modular Java-based framework designed specifically for such applications.

A central principle of scientific computing is that insight emerges through structured visualization. The primary use case for MDI is the researcher employing visualization to interpret complex data, whether originating from simulation, modeling, or real-time data acquisition. Applications range from debugging complex detector systems to communicating experimental results to broader audiences.

The design of MDI is informed by classical principles of modular decomposition \cite{parnas1972criteria}. Rather than organizing functionality around graphical widgets alone, the framework decomposes responsibilities along architectural boundaries: simulation, visualization, messaging, and application coordination. This separation enables scientific applications to evolve over time without entangling rendering logic with computational engines.

\section{Architectural Requirements}

Through extended development of research and instructional tools, several recurring architectural requirements emerged:

\subsection{Multi-Document Support}

Scientific applications often consist of multiple independent yet coordinated views, including plots, simulation windows, diagnostic panels, and control interfaces. A multi-document architecture enables independent lifecycle management while preserving shared infrastructure.

\subsection{Thread-Safe Visualization}

Simulations and data acquisition processes typically execute on background threads. In contrast, the Java Swing graphical toolkit employs a single-threaded rendering model centered on the Event Dispatch Thread (EDT). All UI state mutations and repaint operations must occur on the EDT to ensure correctness and prevent race conditions.

As discussed in the concurrency literature~\cite{goetz2006jcip}, improper coordination between worker threads and the Event Dispatch Thread can lead to inconsistent state, deadlocks, or repaint storms. Consequently, Swing-based systems require disciplined synchronization and strict thread confinement of UI updates.

The MDI framework adopts coalesced update strategies and structured simulation stepping to ensure that background computations interact safely with the EDT. By explicitly separating simulation execution from visualization scheduling, the framework maintains responsiveness while preserving thread safety.

\subsection{Simulation Integration}

A built-in step-based simulation engine supports:

\begin{itemize}
    \item Controlled stepping
    \item Cancellation support
    \item Reset hooks
    \item Coordinated refresh scheduling
\end{itemize}

This ensures deterministic interaction between simulation logic and visualization layers.

\subsection{Dependency Isolation}

Hardware-accelerated 3D rendering via JOGL~\cite{jogl} introduces native and platform-specific dependencies. JOGL is a mature and widely used OpenGL binding for Java. Native libraries introduce potential compatibility risks across evolving JDK releases and operating systems. To preserve lightweight deployment for 2D-only applications, 3D functionality is separated into an optional Maven module. If 3D is not required, then only a dependency on the pure-Java core 2D framework is needed. 

This separation aligns with established principles of modular decomposition~\cite{parnas1972criteria}.

\section{Framework Design}

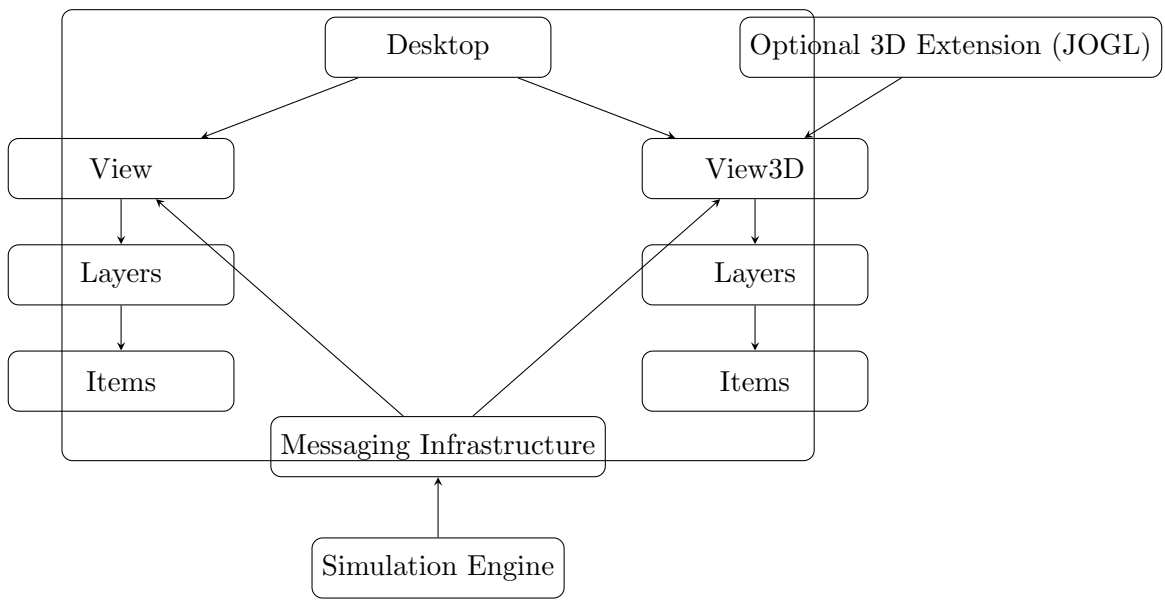
\begin{figure}[H]
\centering
\begin{tikzpicture}[
    box/.style={draw, rounded corners, minimum width=3cm, minimum height=0.8cm, align=center},
    bigbox/.style={draw, rounded corners, minimum width=10cm, minimum height=6cm},
    >=stealth
]

\node[bigbox] (app) {};

\node[box] at (0,2.5) (desktop) {Desktop};

\node[box, below left=0.8cm and 1.2cm of desktop] (view1) {View};
\node[box, below right=0.8cm and 1.2cm of desktop] (view2) {View3D};

\node[box, below=0.6cm of view1] (layers1) {Layers};
\node[box, below=0.6cm of view2] (layers2) {Layers};

\node[box, below=0.6cm of layers1] (items1) {Items};
\node[box, below=0.6cm of layers2] (items2) {Items};

\node[box, below=4.5cm of desktop] (messaging) {Messaging Infrastructure};

\node[box, below=0.8cm of messaging] (simulation) {Simulation Engine};

\node[box, right=2.5cm of desktop] (ext3d) {Optional 3D Extension (JOGL)};

\draw[->] (desktop) -- (view1);
\draw[->] (desktop) -- (view2);

\draw[->] (view1) -- (layers1);
\draw[->] (view2) -- (layers2);

\draw[->] (layers1) -- (items1);
\draw[->] (layers2) -- (items2);

\draw[->] (simulation) -- (messaging);
\draw[->] (messaging) -- (view1);
\draw[->] (messaging) -- (view2);

\draw[->] (ext3d) -- (view2);

\end{tikzpicture}
\caption{Architectural overview of the MDI framework. 
The desktop manages multiple views, each composed of layers and items. 
A messaging infrastructure coordinates state changes, while the simulation 
engine executes background computations. Optional 3D functionality is 
isolated into a separate extension module.}
\label{fig:architecture}
\end{figure}

Figure~\ref{fig:architecture} summarizes the architectural structure of the framework. 
At the highest level, an application consists of a desktop container responsible 
for lifecycle management and coordination of multiple views. Each view is 
hierarchically composed of layers, and each layer contains interactive items. 
This strict containment model clarifies rendering order, input routing, and 
state ownership.

The simulation engine operates independently of the rendering hierarchy. 
Rather than interacting directly with view objects, simulation components 
communicate through a lightweight messaging infrastructure. This indirection 
reduces coupling between computation and visualization, allowing views to 
subscribe selectively to state changes without requiring explicit references 
to simulation internals. Simulation execution occurs on background threads, while visualization updates are scheduled onto the Event Dispatch Thread, enforcing an explicit concurrency boundary within the architecture.

Optional 3D functionality is isolated into a separate module that integrates 
at the view level. As shown in Figure~\ref{fig:architecture}, the 3D extension 
does not alter the core desktop or messaging abstractions, preserving a 
minimal dependency surface for 2D-only applications.

This layered and modular structure reflects established principles of system 
decomposition~\cite{parnas1972criteria}. By constraining communication paths and enforcing explicit ownership boundaries, the framework reduces architectural entropy as applications grow in complexity.

\subsection{View Abstraction}

The core abstraction is the \textit{view}, representing an independent application window integrated into a shared desktop. The desktop may be extended to virtual workspaces, allowing large numbers of views to coexist without overwhelming the physical display.

\subsubsection{Anatomy of a View}

\begin{figure}[H]
\centering
\includegraphics[scale=0.4]{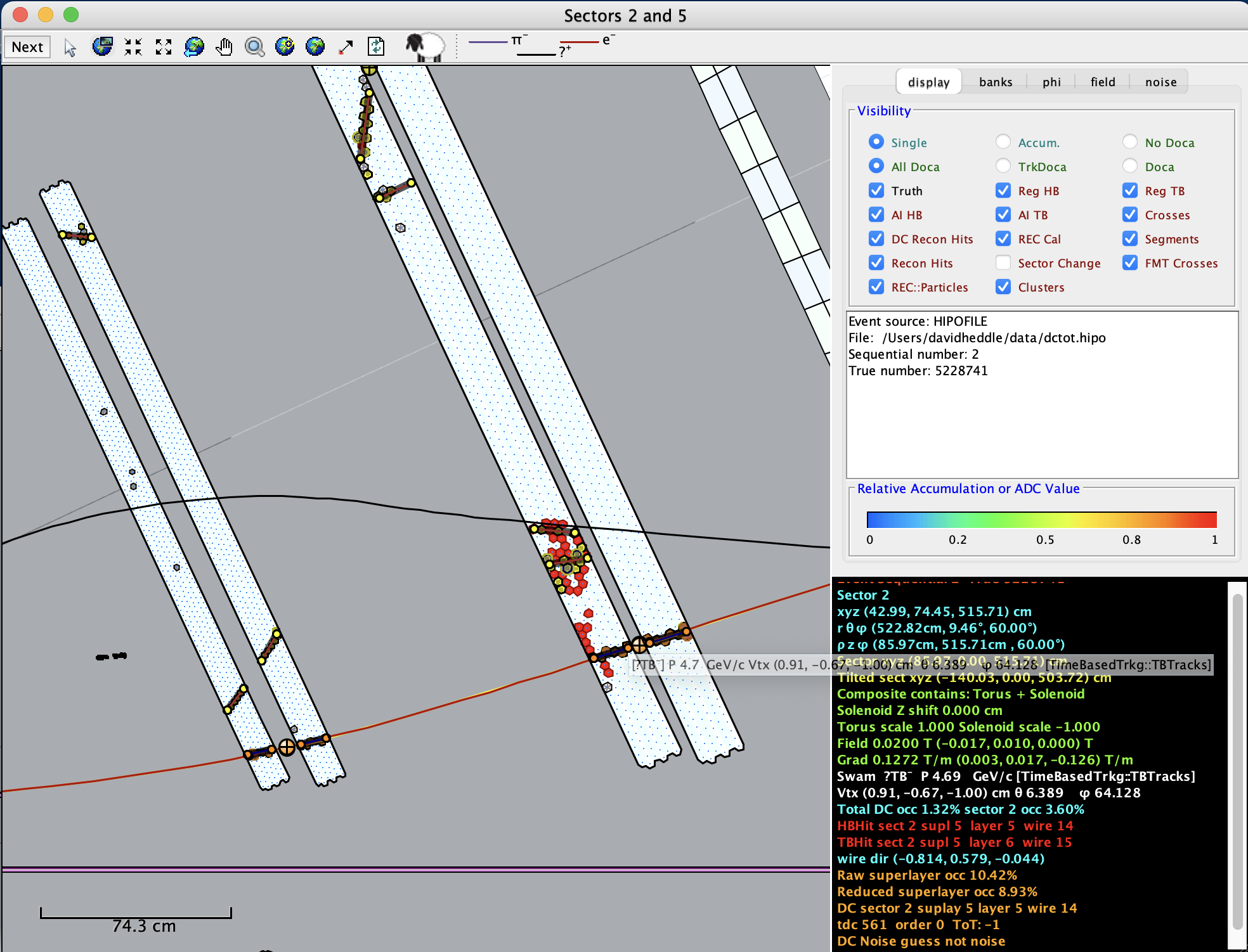}
\caption{A sample view. This view contains components supported by the MDI framework including content, controls, a toolbar, and context-sensitive feedback.}
\label{ced1}
\end{figure}

The optional components of a view are illustrated in Figure~\ref{ced1}. 
This example, drawn from the CLAS12 Event Display~\cite{ziegler} 
and originally developed using the predecessor framework bCNU, 
illustrates the architectural requirements that motivated the design of MDI.

Views may function as quasi-independent sub-applications while retaining the ability to share data efficiently within a single JVM. Their main component is usually graphical, but can be purely text-based or controller-based.

An MDI view provides structured support for the following elements:

\begin{itemize}

\item \textbf{Main content area.} 
The primary rendering surface responsible for visualization. 
In this example it displays detector geometry, reconstructed particle tracks, 
and associated signal information. The content area is layered, 
allowing independent rendering and interaction of geometric, 
reconstruction, and annotation elements.

\item \textbf{Extensible toolbar.} 
A configurable toolbar supporting interaction tools such as zooming, 
panning, and navigation. Tool configuration is declarative and 
independent of the rendering implementation.

\item \textbf{Control panel.} 
An auxiliary interface region used for configuration and state management. 
In this example, the tabbed panel in the upper right controls visibility, 
accumulation modes, and display parameters. Such panels operate independently 
of the rendering surface while communicating through a lightweight messaging model.

\item \textbf{Feedback panel.} 
A diagnostic region that presents contextual system state. 
Here, the lower-right panel provides event metadata and field values, 
updated in response to user interaction.

\item \textbf{Hover feedback.} 
Transient context-sensitive overlays associated with rendered items. 
This mechanism provides detailed inspection without cluttering the 
persistent visualization.

\end{itemize}

The coexistence of these components within a single view illustrates the need 
for clear separation of rendering layers, input handling, simulation state, 
and user interface controls. These requirements directly informed the 
modular abstractions adopted in the MDI framework.

\subsection{Layered Rendering Model}

Rendering is organized into Z-ordered \textit{layers} containing interactive \textit{items}. This enables:

\begin{itemize}
    \item Structured drawing separation
    \item Selective repainting
    \item Clear input event routing
\end{itemize}

The user can define new layers, reorder layers (with a caveat discussed below), and toggle layer visibility.

A design constraint is that two predefined layers are reserved: the connection layer and the annotation layer. These special layers cannot be moved or deleted. The connection layer is always drawn first, so that connections are drawn underneath all items.  Annotations are drawn last, so they always appear on top. 

\subsection{Items}

So far we have seen that a MDI application is comprised of views, and views are comprised of layers. The terminal abstraction in this hierarchy is the \textit{item}. An item typically represents something ``real", like a nuclear physics detector or a network node. Items can be dragged, rotated, selected, resized, locked, deleted, styled, and edited. All such capabilities are available but optional.

There is a set of base classes that the user can extend:
\begin{itemize}
    \item ConnectorItem. A connector between two items
    \item EllipseItem. An oval shaped item.
    \item ImageItem. Displays an image.
    \item LineItem. A line connecting two points.
    \item PointItem. A point drawn with a glyph.
    \item PolygonItem. A closed polygon.
    \item PolylineItem. An open polygon.
    \item RadArcItem. A pie-slice ``wedge'' shape.
    \item RectangleItem. A rectangular shaped item.
    \item TextItem. A multi-line text item.
\end{itemize}

\begin{figure}[H]
\centering
\includegraphics[scale=0.4]{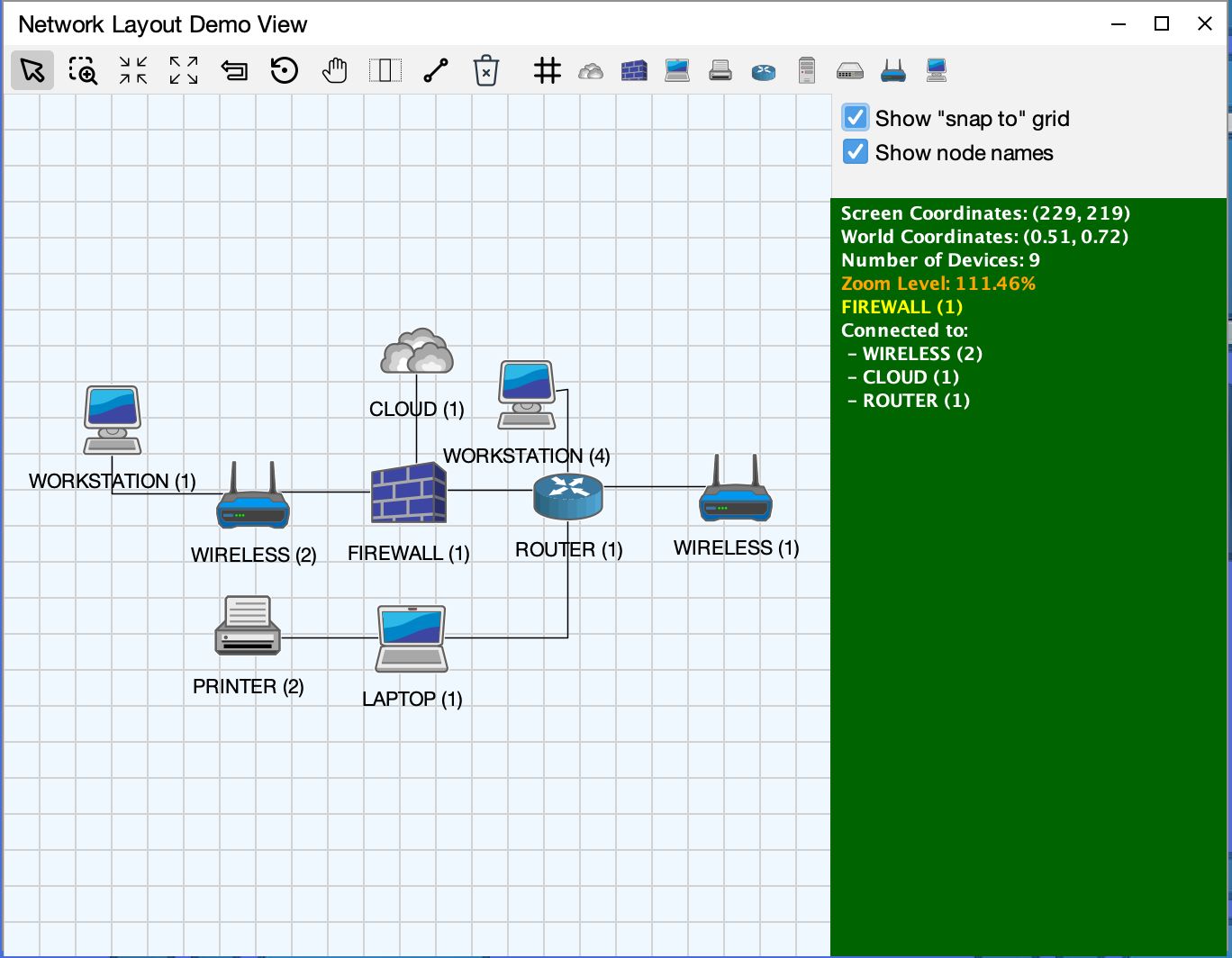}
\caption{An example of a view with user-defined items. The network nodes extend the base class RectangleItem, from which they inherit a rich set of capabilities including dragging, editing, responding to toolbar tools, reordering and more.}
\label{layout}
\end{figure}

Complex items are created by extending one of the base classes. Figure~\ref{layout} shows a very simple view with network nodes that extend the base class RectangleItem, adding the ability to display an SVG icon and to draw a string (the device name) underneath the icon.

\subsection{Event Coordination}

A lightweight messaging infrastructure coordinates state changes across views without tight coupling. The messaging system enables loosely coupled interaction between views, 
control panels, and simulation components. Messages are dispatched 
asynchronously and may be filtered or scoped to specific views. 
This design avoids direct object references across subsystems, 
reducing architectural rigidity and improving extensibility.

\subsection{Simulation Engine}

The simulation engine operates as a deterministic step loop executed on background threads. After each simulation step, update events are coalesced and dispatched to interested views. By decoupling computation from rendering frequency, the framework avoids repaint storms and preserves interactive responsiveness even during intensive numerical updates.

Simulation state transitions are explicit and observable, enabling controlled stepping, 
reset, and cancellation semantics. This structured model avoids ad hoc thread creation 
within views and centralizes concurrency management.

The user configures the engine threading through the timing parameters:
\begin{itemize}
    \item Refresh interval: Target interval for posting refresh events to the EDT.
    \item Progress interval: Target interval for posting progress ``ping'' events to the EDT.
    \item Cooperative yield: Rate limit for the simulation thread. It is not interpreted as ``sleep this long each step''. It is a minimum interval between opportunities (which may comprise many simulation steps) for the simulation to yield CPU time. 
    
\end{itemize}

The engine never directly mutates Swing components; instead, it posts structured update messages to the EDT.
\subsection{Integrated Plotting}

\begin{figure}[H]
\centering
\begin{subfigure}{0.48\textwidth}
  \centering
  \includegraphics[width=\linewidth]{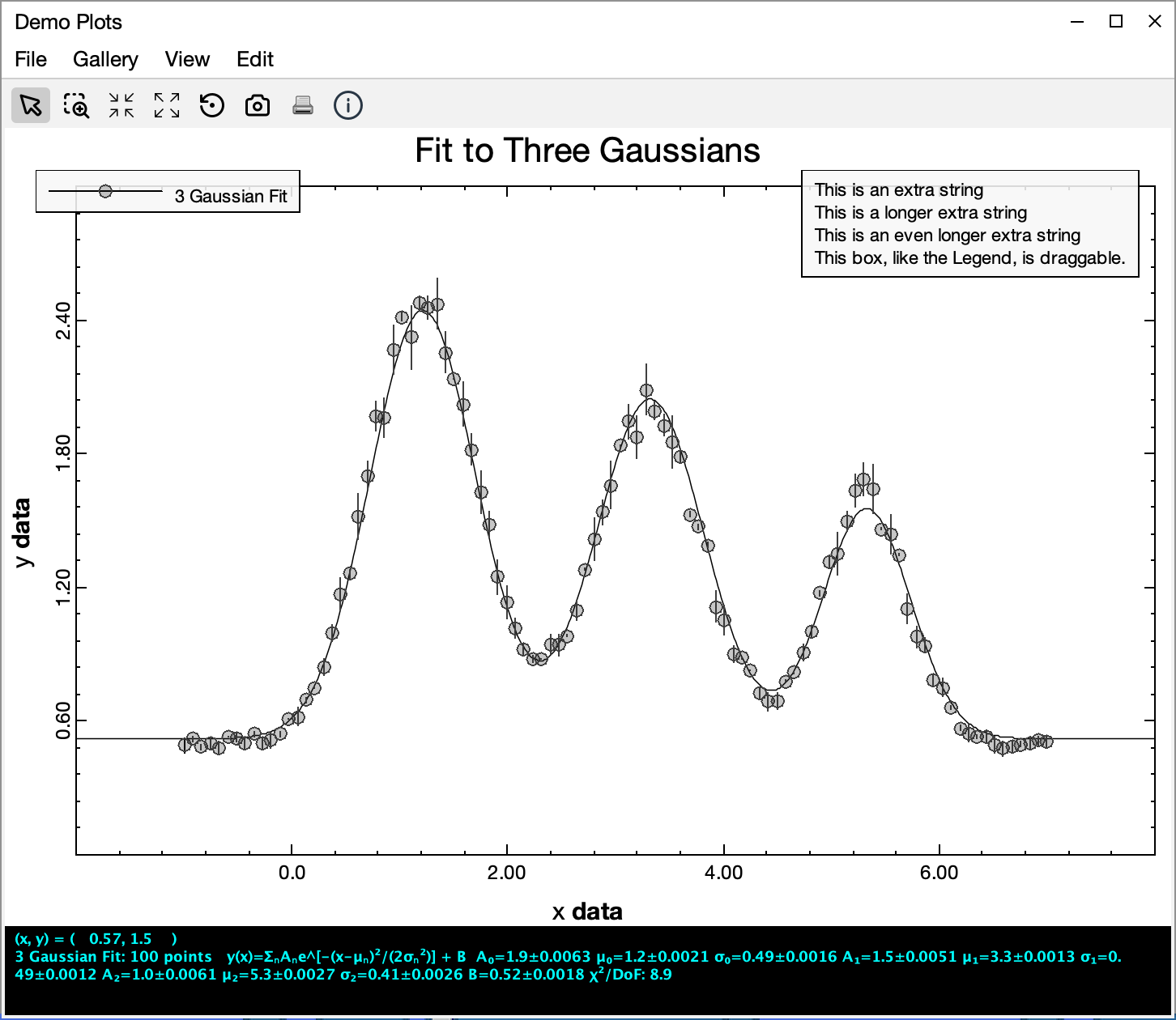}
  \caption{A triple Gaussian fit.}
\end{subfigure}
\hfill
\begin{subfigure}{0.48\textwidth}
  \centering
  \includegraphics[width=\linewidth]{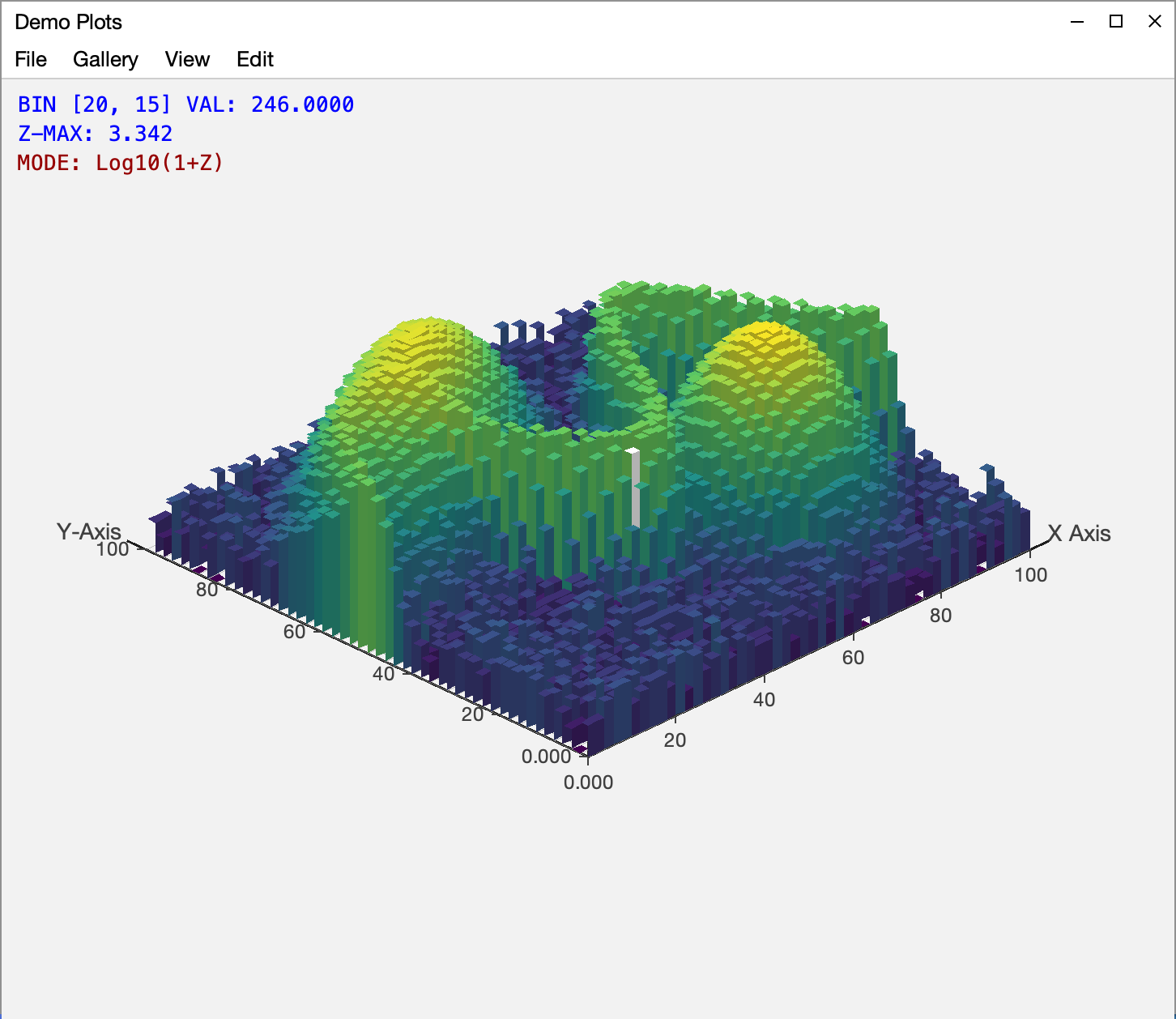}
  \caption{A 2D Histogram}
\end{subfigure}
\caption{The integrated sPlot package provides a variety of plot types and curve fitting options.}
\label{fig:plots}
\end{figure}

MDI comes with sPlot, an integrated plotting package. Plots are interactive, editable, and can be persisted. An extensible collection of curve-fitting options are provided through a dependency on the Apache Commons Math library\cite{commonsmath}. Two examples are shown in Figure~\ref{fig:plots}.

The plotting subsystem adheres to the same layered and message-driven design as other views, allowing plots to respond to simulation events without direct coupling to simulation internals.

\FloatBarrier

\section{Optional 3D Extension Module}

3D functionality is implemented in a separate Maven artifact to avoid coupling the core framework to JOGL. This design provides:

\begin{itemize}
    \item Reduced dependency surface for 2D applications
    \item Improved long-term JVM compatibility
    \item Explicit architectural opt-in for OpenGL support
\end{itemize}

Both 2D and 3D views may coexist within a single application instance.

\section{Case Study: Gas Expansion Simulation}

\begin{figure}[H]
\centering
\begin{subfigure}{0.48\textwidth}
  \centering
  \includegraphics[width=\linewidth]{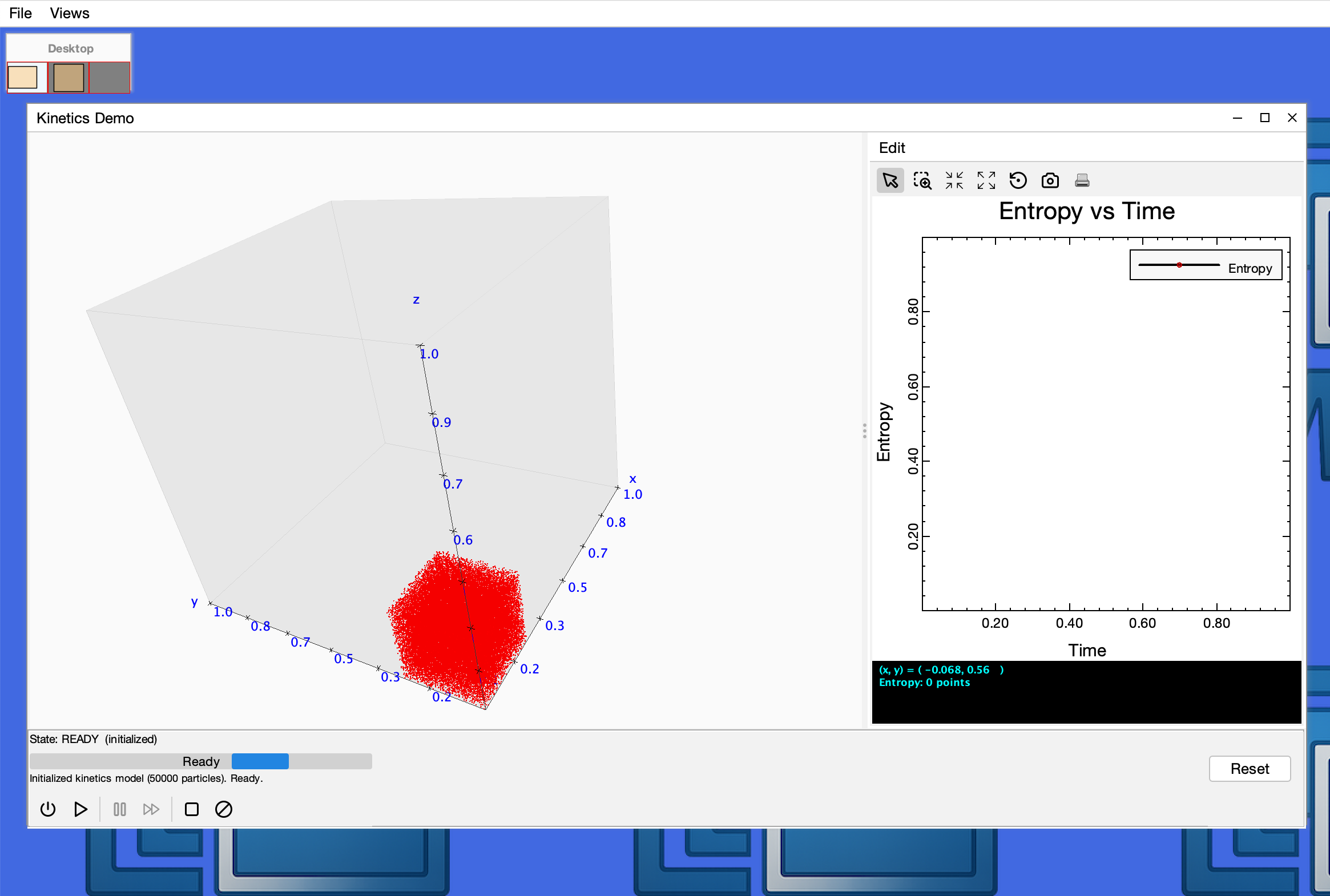}
  \caption{Initial state: particles confined to a corner.}
\end{subfigure}
\hfill
\begin{subfigure}{0.48\textwidth}
  \centering
  \includegraphics[width=\linewidth]{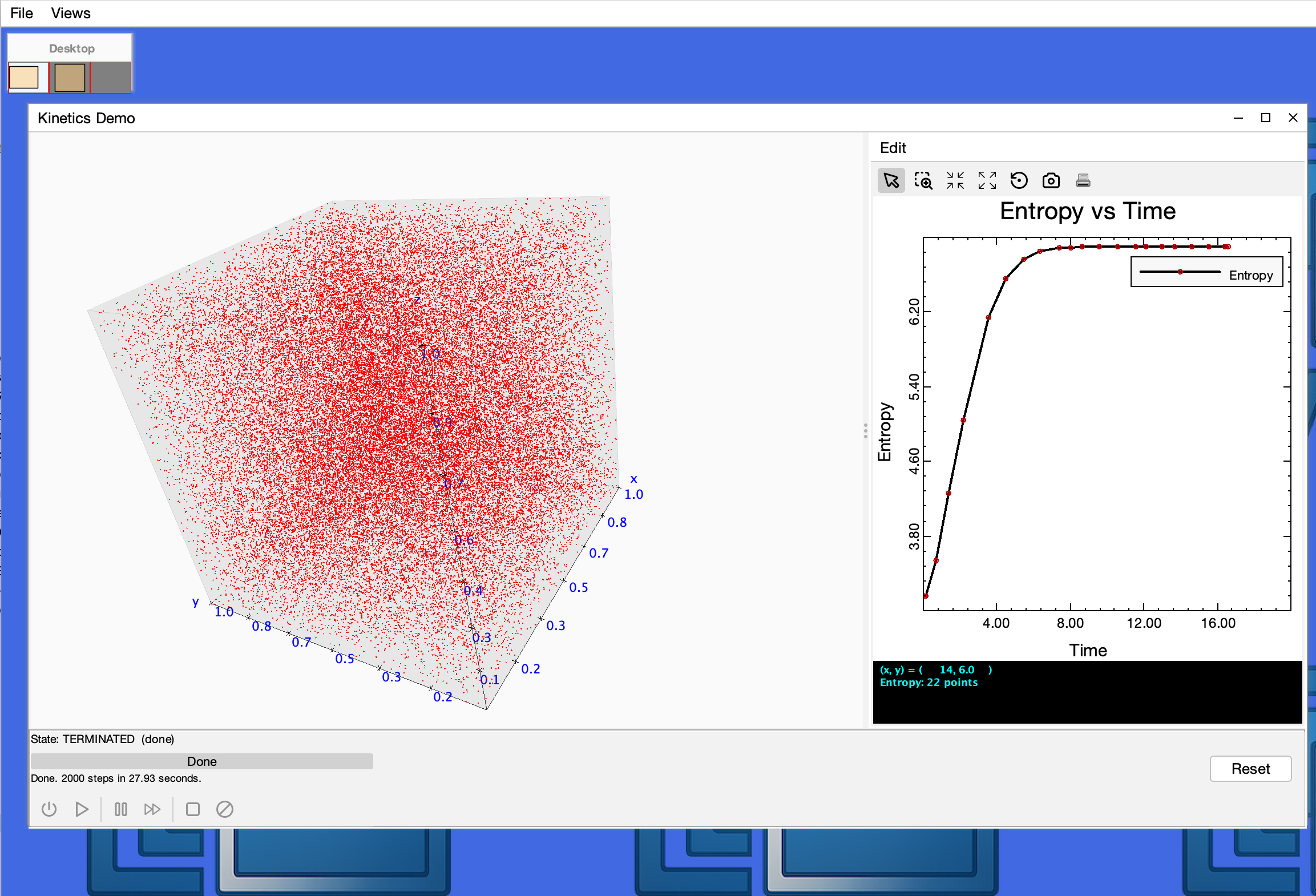}
  \caption{Final state: particles uniformly distributed.}
\end{subfigure}
\caption{3D gas expansion simulation with synchronized entropy plot. The 3D view renders particle motion while a 2D plot tracks entropy over time.}
\label{fig:kinetics}
\end{figure}

The 3D gas expansion example that tracks the free expansion of 50,000 particles in a 3D view (see Figure~\ref{fig:kinetics}) serves as a compact integratin demonstration of the framework architecture. The particle simulation executes on background threads using a deterministic stepping model, while rendering is handled by the optional OpenGL extension module. Simultaneously, a synchronized 2D plot tracks entropy as a function of time, illustrating coordinated multi-view updates.

The demonstration integrates a real-time 3D particle simulation, synchronized 2D entropy plotting, and interactive control panels for simulation management.

This example also demonstrates that the messaging infrastructure, simulation engine, and layered rendering model operate cohesively across both 2D and 3D components. Importantly, the 3D functionality is entirely optional; the same architectural structure supports purely 2D applications without modification.
\FloatBarrier

\section{Discussion}

The framework prioritizes stability and modularity over contemporary UI trends. While alternatives such as JavaFX and web-based frameworks offer modern interface paradigms, they introduce trade-offs in dependency footprint, deployment complexity, or long-term ecosystem uncertainty.

The design presented here emphasizes architectural clarity, dependency isolation, and deterministic behavior suitable for scientific and engineering contexts.

The framework does not attempt to compete with modern declarative UI systems 
in terms of visual styling or rapid prototyping workflows. Instead, it targets 
applications where long-term maintainability, deterministic execution, and 
tight integration between computation and visualization are primary concerns.

A potential limitation is that the framework remains within the Swing ecosystem, 
which, while stable, is no longer the focus of active feature expansion. 
However, this stability is viewed here as an asset for scientific applications 
with multi-decade lifecycles.

Unlike frameworks that tightly couple rendering, simulation, and interface logic, MDI enforces explicit boundaries between these concerns. This separation improves maintainability, testability, and long-term extensibility—qualities that are particularly important for scientific software with multi-year lifecycles.

\section{Availability}

The framework is publicly available:

\begin{itemize}
    \item GitHub: \url{https://github.com/heddle/mdi}
    \item Maven Central: \texttt{io.github.heddle:mdi}
\end{itemize}

To build and run the demo application from source:
\begin{verbatim}
mvn clean package
mvn exec:java -Dexec.mainClass="edu.cnu.mdi.demo.DemoApp"
\end{verbatim}

The 3D extension is also publicly available:
\begin{itemize}
    \item GitHub: \url{https://github.com/heddle/mdi-3D}
    \item Maven Central: \texttt{io.github.heddle:mdi-3D}
\end{itemize}

To build and run the 3D demo application from source:

\begin{verbatim}
mvn clean package
mvn exec:java -Dexec.mainClass="edu.cnu.mdi3D.app.DemoApp3D"
\end{verbatim}
You can directly import MDI and MDI-3D into your IDE as Maven projects.  MDI and MDI-3D are distributed under the MIT License.

\section{Conclusion}

Long-lived scientific desktop applications require architectural discipline distinct from that emphasized in rapidly evolving UI ecosystems. By prioritizing modularity, thread safety, and dependency isolation, the presented framework provides a stable foundation for visualization and simulation in the JVM environment.

The framework is released as open-source software under the MIT license and is intended to support reproducible scientific workflows. By providing architectural infrastructure rather than domain-specific tooling, MDI offers a stable foundation for long-lived scientific desktop applications requiring integrated visualization and simulation. The design principles described here may also inform other modular JVM-based desktop systems.

\bibliographystyle{plain}
\bibliography{references}

\end{document}